\documentclass[twocolumn,showpacs,superscriptaddress,prl]{revtex4}
\usepackage{amsbsy,amssymb,amsmath,bm}
\usepackage{graphicx,color}

\newcommand{\bp}{\mathbf{p}}

\renewcommand{\Im} {\mathop{\mathrm{Im}}}

\begin{document}

\title{Charge transfer between a superconductor and a hopping insulator}
\author{V. I. Kozub}
\affiliation{A. F. Ioffe Physico-Technical Institute of Russian Academy of Sciences,
194021 St. Petersburg, Russia}
\affiliation{Argonne National Laboratory, 9700 S. Cass Av., Argonne, IL 60439, USA}
\author{A. A. Zyuzin}
\affiliation{A. F. Ioffe Physico-Technical Institute of Russian Academy of Sciences,
194021 St. Petersburg, Russia}
\author{Y. M. Galperin}
\affiliation{Department of Physics and Center for Advanced Materials and Nanotechnology,
University of Oslo, PO Box 1048 Blindern, 0316 Oslo, Norway}
\affiliation{A. F. Ioffe Physico-Technical Institute of Russian Academy of Sciences,
194021 St. Petersburg, Russia}
\affiliation{Argonne National Laboratory, 9700 S. Cass Av., Argonne, IL 60439, USA}
\author{V. Vinokur}
\affiliation{Argonne National Laboratory, 9700 S. Cass Av., Argonne, IL 60439, USA}

\begin{abstract}
We develop a theory of the low-temperature charge transfer between a
superconductor and a hopping insulator. We show that the charge transfer is
governed by the coherent two-electron -- Cooper pair conversion process,
\textit{time reversal reflection}, where electrons tunnel into
superconductor from the localized states in the hopping insulator located
near the interface, and calculate the corresponding interface resistance.
This process is an analog to conventional Andreev reflection process. We
show that the time reversal interface resistance is accessible
experimentally, and that in mesoscopic structures it can exceed the bulk
hopping resistance.
\end{abstract}

\date{\today}
\maketitle

The transmission of the charge through the normal metal-superconductor
interface occurs via the electron-hole conversion known as the Andreev
reflection process: an electron incident from the metal side with an energy
smaller than the energy gap in the superconductor is converted into a hole
which moves backward with respect to the electron. The missing charge $2e$
(an electron has charge $-e$ and a hole $+e$) propagates as an electron pair
into the superconductor and joins the Cooper pair condensate~\cite{Andreev}.
Correspondingly, a Cooper pair transfer from the superconductor is described
as the Andreev reflection of a hole. This Andreev transport channel is
characterized by the so-called Andreev interface contact resistance. Since
transport current is introduced into a superconductor via normal leads, the
Andreev reflection phenomenon is a foundation for most applications of
superconductors (see Ref.~\cite{Book} for a review).

There exists however an important experimental situation of the hopping
insulator coupled to a measuring circuit via superconducting leads (see, for
example,~\cite{Rentch}), where the conventional Andreev reflection picture
does not apply. The transport in hopping semiconductors occurs via localized
(\textit{non-propagating}) single particle states~\cite{end1} with undefined
momentum and therefore a Cooper pair on the superconductor side cannot form.
A single particle transport through the interface is exponentially
suppressed, $\propto e^ {-\Delta/T}$, where $\Delta$ is the superconductor
gap, the temperature, $T$, being measured in energy units; therefore to
explain the finite conductivity observed in experiments one needs
Andreev-type processes capable to facilitate two particle transfer through
the hopping insulator/superconductor interface allowing for Cooper pair
formation. The possibility of such a transfer through the
hopping-superconductor interface was discussed in Ref.~\cite{Agrinskaya} but
no quantitative theory of hopping transport - supercurrent conversion was
presented.

In this Letter we develop a theory for the transport through the hopping
insulator-superconductor interface and derive the corresponding contact
resistance. We show that the low-temperature charge transfer occurs via the
correlated processes mediated by the \textit{pairs} of hopping centers
located near the interface. We demonstrate that this process resembles the
conventional Andreev electron-to-hole reflection into a normal metal, the
exponential suppression of transport specific to a single-particle processes
being lifted. Thus, despite the limitation in the number of coherent hopping
centers that can carry Andreev transport, the resulting contact resistance
can become low as compared to the resistance of the hopping insulator.
However in mesoscopic structures the interface resistance can be comparable
or even exceed the hopping resistance. 
The proposed mechanism resembles the so-called crossed Andreev charge
transfer~\cite{CA}, discussed recently in connection with a
superconductor-dot entangler~\cite{ent,ent1}. The difference is that in~\cite%
{CA,ent,ent1} the transport mediated by artificial quantum dots was
considered. In our case, the transport occurs via randomly located sites in
the hopping insulator (HI), and the main problem one has to solve is finding
the optimal configuration of the sites responsible for the charge transfer.
Hereafter we will refer to the proposed charge transfer mechanism as to the
\textit{time reversal reflection}.

\begin{figure}[h]
\centerline{
\includegraphics[width=5cm]{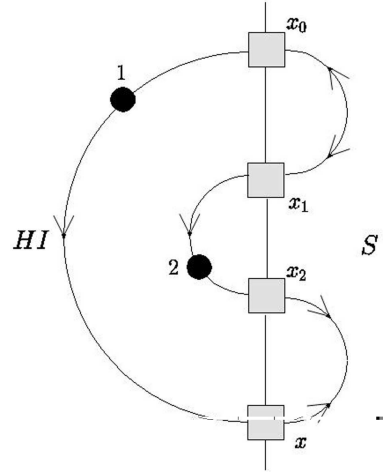}
}
\caption{The diagram describing the time reversal reflection. Lines with one
arrow correspond to the Green's functions in the hopping insulator. They are
associated either with the center 1, or with the center 2. Lines with two
arrows correspond to anomalous Green's functions, see {\protect\cite{AGD}}.
Squares correspond to matrix elements of the tunneling Hamiltonian~{(\protect
\ref{TH})}. }
\label{fig:1}
\end{figure}

Let a superconductor (S) and an HI to occupy the adjacent 3D semi-spaces
separated by a tunneling barrier (B). The presence of the barrier simplifies
calculations which will be made in the lowest non-vanishing approximation in
the tunneling amplitude $T_{0}$. This models the Schottky barrier usually
presenting at a semiconductor-metal interface. In the linear response theory
the conductance is determined by the Kubo formula~\cite{Kubo} for the
susceptibility,
\begin{equation}
\chi (\omega )=i\int\limits_{0}^{\infty }\left\langle \left[ \hat{I}^{+}(t),%
\hat{I}(0)\right] \right\rangle e^{i\omega t}\,dt  \label{in}
\end{equation}%
as $\mathcal{G}=\lim_{\omega \rightarrow 0}\omega ^{-1}\Im \chi (\omega )$.
Here the current operator $\hat{I}(t)$ is defined as~\cite{tunnel-current}:
\begin{equation*}
\hat{I}(t)=ied\,T_{0}\int d^{2}r\,[a^{+}(\mathbf{r},t)b(\mathbf{r},t)-\text{%
h.c.}]\,,
\end{equation*}%
where $\mathbf{r}$ is the coordinate in the interface plane, $a^{+}(\mathbf{r%
},t)$ and $b(\mathbf{r},t)$ are creation and annihilation operators in the
semiconductor and superconductor, respectively, $d$ is the electron
localization length under barrier. The susceptibility, $\chi (\omega )$, is
calculated by analytical continuation of the Matsubara susceptibility~\cite%
{AGD},
\begin{equation}
\chi _{M}(\Omega )=\int_{0}^{\beta }\left\langle T_{\tau }I(\tau
)I(0)\right\rangle e^{i\Omega \tau }\,d\tau \,.
\end{equation}%
Here $T_{\tau }$ means ordering in the imaginary time, $\beta \equiv 1/T$.
In the expression for $\left\langle T_{\tau }I(\tau )I(0)\right\rangle $ one
should expand to the second order with respect to the tunneling Hamiltonian,
\begin{equation}
H_{T}(\tau )=dT_{0}\int d^{2}r\left[ a^{+}(\mathbf{r},\tau )b(\mathbf{r}%
,\tau )+\text{h.c.}\right] \,.  \label{TH}
\end{equation}%
Keeping only those second order terms that contain $\left\langle T_{\tau }b(%
\mathbf{r},\tau )b(\mathbf{r}_{0},0)\right\rangle \left\langle T_{\tau
}b^{+}(\mathbf{r}_{1},\tau _{1})b^{+}(\mathbf{r}_{2},\tau _{2})\right\rangle
$ products and thus represent the time reversal scattering which we are
interested in, one arrives at the expression
\begin{eqnarray}
&&\left\langle T_{\tau }\hat{I}(\tau )\hat{I}(0)\right\rangle
=e^{2}|T_{0}|^{4}\int d\tau _{1}\,d\tau _{2}\prod_{i}d^{2}r_{i}(A+B)\,;
\notag \\
&&A(\{x_{i}\})=F(x-x_{0})F^{+}(x_{1}-x_{2})G(x_{1},x)G(x_{2},x_{0})\,,
\notag \\
&&B(\{x_{i}\})=F(x-x_{1})F^{+}(x_{0}-x_{2})[G(x_{0},x)G(x_{2},x_{1})  \notag
\\
&&\qquad \qquad \qquad -G(x_{0},x_{1})G(x_{2},x)]\,,  \label{gf1}
\end{eqnarray}%
where $x\equiv \{\mathbf{r},\tau \}$, $x_{0}\equiv \{\mathbf{r}_{0},0\}$, $%
x_{i}\equiv \{\mathbf{r}_{i},\tau _{i}\}$; $F(x-x^{\prime })=\langle T_{\tau
}b(\mathbf{r},\tau )b(\mathbf{r}^{\prime },\tau ^{\prime })\rangle $ is the
anomalous Green function in the superconductor while $G(x,x^{\prime
})=-\langle T_{\tau }a(\mathbf{r},\tau )a^{+}(\mathbf{r}^{\prime },\tau
^{\prime })\rangle $ is the Green function in the hopping insulator. One can
show that the Andreev-type process we are interested in is given by the
first term of $B(\{x_{i}\})$ in Eq.~(\ref{gf1}). The relevant diagram is
shown in Fig.~\ref{fig:1}.
Keeping only this term and using the Matsubara frequency representation one
obtains
\begin{eqnarray}
&&\chi _{M}(\Omega )=2Te^{2}|T_{0}|^{4}d^{4}\int
\prod_{i}d^{2}r_{i}\sum_{\omega _{n}}F(\mathbf{r}-\mathbf{r}_{1},\omega _{n})
\notag  \label{1} \\
&&\times F^{+}(\mathbf{r}_{0}-\mathbf{r}_{2},\omega _{n})G(\mathbf{r}_{0},%
\mathbf{r},\omega _{n}-\Omega _{m})G(\mathbf{r}_{2},\mathbf{r}_{1},-\omega
_{n}),
\end{eqnarray}%
where $\Omega _{m}=2\pi mT$ and $\omega _{n}=(2n+1)\pi T$. The normal
Green's functions can be expressed through the wave functions of the
localized states, $\varphi _{s}(\mathbf{r})=(\pi a^{3})^{-1/2}\exp (-|%
\mathbf{r}-\mathbf{r}_{s}|/a)$, as
\begin{equation}
G(\mathbf{r},\mathbf{r}^{\prime },\omega _{n})=\sum_{s}\frac{\varphi
_{s}^{\ast }(\mathbf{r})\varphi _{s}(\mathbf{r}^{\prime })}{i\omega
_{n}-\varepsilon _{s}}\,. \\
\end{equation}%
We have assumed that for all of the sites under consideration the voltage
drops between the site and the superconductor are the same. This is true
when the partial interface resistance due to a time reversal pair is much
larger than the typical resistance of the bond forming the percolation
cluster. This situation resembles that considered by Larkin and Shklovskii
for the tunnel resistance between the hopping conductors \cite{Larkin}.

The anomalous Green function, $F(R,\omega_n)$, is
\begin{eqnarray}
F(R,\omega_n)&=&\int \frac{d^{3}p}{(2\pi \hbar)^3}\frac {\Delta}{%
\Delta^{2}+\xi^{2}_\bp +\omega_n^{2}}\, e^{i\mathbf{p} \cdot \mathbf{R}%
/\hbar}  \notag \\
&=& \frac{\pi g_{m} \Delta}{2\sqrt{\Delta^{2}+\omega_n^{2}}} \frac{%
\sin(Rk_{F})}{Rk_{F}} e^{-\frac{R}{\pi\xi}\frac{\sqrt {\Delta^{2}+%
\omega_n^{2}}}{\Delta}}\, .
\end{eqnarray}
Here $\xi_\bp=(p^{2}-p_F^2)/2m$, $g_{m}=mp_{F}/\pi^{2}\hbar^3$ is the
density of states in a metal, $k_F=p_F/\hbar$, while $\xi$ is the coherence
length in a superconductor. Since $F(R)$ oscillates with the period $2\pi/k_F
$ integration over spatial coordinates along the interface yields the factor
$a^4/k_F^6|\bm{\rho}_{ls}|^2$. Here $\bm{\rho}_{ls}$ is projection of the
vector $\mathbf{R}_{ls}$ connecting the centers on the interface plane.
Note that the dependencies on $R$ and $\xi$ are similar to that given by
Eq.~(21) of the paper \cite{ent} for the pair of the quantum dots near the
superconducting interface. However the latter equation does not specify the
dependences of the transmission coefficients on the real physical parameters
of the interface and the localized centers.

The summation over the Matsubara frequencies, $\omega_n$, is standard,
\begin{equation*}
T\sum_{\omega_n}f(\omega_n) =\oint \frac{d \varepsilon}{4\pi i}\,
f(\varepsilon)\tanh \frac{\varepsilon}{2T}\, .
\end{equation*}
The contour of integration closes the cuts $|\varepsilon | >\Delta$ along
the real axis. Upon analytical continuation one arrives at the following
expression for the conductance:
\begin{eqnarray}  \label{3}
&&\mathcal{G} =\frac{\pi e^2 g_{m}^{2} |T_{0}|^{4}d^{4}}{2\hbar Tk_F^6 a^2}
\sum_{s \ne l} \frac{n(\varepsilon_s)n(\varepsilon_l)\Delta^2}{|\bm{\rho}%
_{ls}|^2(\Delta^2- \varepsilon_{s}^{2})} e^{-2(z_{s}+z_{l})/a}  \notag \\
&&\times e^{-2|\bm{\rho}_{ls} |\sqrt{\Delta^{2}-\varepsilon_{s}^{2}}/\pi\xi
\Delta} \delta \left(\varepsilon_{s}+\varepsilon_{l}+U_c\right)\, .
\end{eqnarray}
Here $n(\varepsilon) \equiv \left(e^{\varepsilon/T}+1\right)^{-1}$ is the
Fermi distribution, $U_c$ is the energy of the inter-site Coulomb repulsion,
and $z$-axis is perpendicular to the interface.

In what follows we will replace $\sum_{l,s}$ by $g^2 \int d^3 r_l\, d^3 r_s
\, d\varepsilon_l \, d \varepsilon_s$ where $g$ is the the effective density
of states in the hopping insulator. This is the density of states in the
layer adjacent to the interface. Due to screening by the superconductor it
is not affected by the Coulomb gap and can be considered as constant. Since
we are dealing with the pairs close to the interface, the Coulomb repulsion
is suppressed by screening. This screening can be conveniently regarded as
an interaction of the charged particle with its image having the opposite
charge. Thus the Coulomb correlation manifest themselves as the
dipole-dipole interaction and for $\rho_{sl} \gg a$ one arrives at $%
U_c=e^2a^2/\kappa \rho_{sl}^3$. Requiring it to me smaller than $T$ one
obtains a cut-off  $\rho_{sl} \ge \rho_T \equiv a (e^2/\kappa a T)^{1/3}$.

As a crude estimate, we take $d^4 \sim k_F^{-4}$, while $T_0 \approx T_p
e^{-\Lambda}$ with $T_p \sim \varepsilon_F$. Bearing this in mind one finds $%
g_m^2 T_p^2 /k_F^6 \sim g_m^2 \varepsilon_F^2 /k_F^6 \sim 1$. Since the
ratio $T_p/(ak_F)^2$ is of the order of the typical energy of the localized
state, $\varepsilon_d \sim \hbar^2/ma^2$, one arrives at the estimate
\begin{equation}  \label{es}
\frac{\mathcal{G}}{\mathcal{G}_n}\sim ga \rho_T^2\varepsilon_d e^{-2
\Lambda}, \quad \mathcal{G}_n \sim \frac{e^2}{\hbar} \, g aS \varepsilon_d\,
e^{-2 \Lambda}.
\end{equation}
Here $\mathcal{G}_n$ is the conductance of a boundary between a normal metal
and a hopping insulator, while $S$ is the contact area. The product $%
gaS\varepsilon_d$ is nothing but the number of localized centers within the
layer of a thickness $a$ near the interface and the factor $ga \rho_T^2
\varepsilon_d$ expresses the probability of finding a critical pair, i.~e.,
a pair of nearly located hopping centers that dominates the time reversal
reflection processes discussed above.

The above approach holds, as we have already mentioned, only if the
resistance of the typical time reversal resistor (TRR) is much larger then
that of the critical hopping resistor, $R_h =(h/e^2 \gamma)\,e^{\zeta}$,
where $\gamma$ is a dimensionless factor depending on the mechanism of
electron-phonon interaction and $\zeta$ is the hopping exponent~\cite%
{Shklovskii-Efros}, i.~e., with the exponential accuracy, as long as $%
4\Lambda > \zeta$.

There are many realistic situations where the barrier strength, $\Lambda$,
is not too large; the Schottky barrier at the natural interface~\cite%
{Agrinskaya} is certainly the case like that. Consequently, if $\zeta \gg 1$%
, i.~e., if the system is far from the metal-to-insulator transition point,
the procedure of summation over the localized states should be modified.
Namely, the choice of the pairs facilitating the charge transfer depends on
the structure of the bonds connecting critical pairs to the rest of
percolation cluster.

According to the above considerations the voltage drops mainly on the bond
connecting percolation cluster in the HI the critical TRR for which the
distance between its pair components is less than the correlation length, $%
\mathcal{L}$, of the backbone cluster. The incoherent electron transport can
be ensured by a single \textit{one} bond connecting the cluster to any of
the TRR sites. Thus, the ratio $\mathcal{G}/\mathcal{G}_n$ is the
probability to find a TRR contacting the percolation cluster.

To estimate this probability let us consider the layer with the thickness of
the typical hopping distance, $r_h$, near the interface where all the bonds
of the backbone cluster have necessarily have a site within this layer. The
total number of states in this layer is $gSr_h \varepsilon_h$ where $%
\varepsilon_h=T\zeta$ is the width of the hopping band. This product can be
estimated as $(\beta/8)(S/r_h)^2$, where $\beta$ is a numerical constant~%
\cite{Shklovskii-Efros}. For the case of Mott variable range hopping (VRH), $%
\beta \approx 20$.

The number of TRRs in this layer can be estimated as follows. Let us note
first that the conserving energy, $\varepsilon_s+ \varepsilon_l+U_c$, from $%
\delta$-function in Eq.~(\ref{3}) is associated with the band given by the
broadening $\nu =\nu_0 \exp(-2 r_d/a)$. Here $r_d$ is distance to the
nearest neighbor in HI. Indeed, 
the most natural source for the broadening of the resonance is coupling of
the localized states. Secondly, since both electrons escape from the TRR
through a single bond, the in-plane distance, $\rho_{sl}$ should not exceed
the typical distance between the hopping sites, $r_h =a\zeta/2 \ll \xi$.
Keeping the exponential accuracy we arrive at the following criterion that
the resistance of TRR is less than the resistance of a typical hopping
resistor:
\begin{equation}  \label{bind}
4 \Lambda +\ln \frac{T}{\nu_0}+\frac{\max(|\varepsilon_s|,|\varepsilon_l|)}{T%
} + \frac{2r_d}{a} + \frac{ 2(z_s + z_l)}{a}< \zeta\, .
\end{equation}
One may consider this equation as a generalization of the ``connectivity
criterion" to include the TRR. Here we deal with the independent variables $%
\varepsilon_s, \varepsilon_l, r_d, z_s$ and $z_l$ over which the averaging
procedure should be done with an account of the restriction of Eq.~(\ref%
{bind}), Thus the number of the relevant TRRs is
\begin{eqnarray}
&&8\pi^2 g^3S \int \mathrm{d}\varepsilon_l\int \! \mathrm{d} \varepsilon_s
\int^{\varepsilon_s}_0 \! \! \mathrm{d}\varepsilon_d \int \! r_d^2\, \mathrm{%
d}r_d \int\! \mathrm{d} z_s \mathrm{d} z_l \int_0^{r_h} \! \! \rho\, \mathrm{%
d} \rho  \notag \\
&& \qquad \times \Theta \left[\frac{\max(\varepsilon_s,\varepsilon_l)}{T} +
\frac{2r_d}{a} + \frac{ 2(z_s + z_l)}{a}- \alpha \zeta\right]  \label{NAR}
\end{eqnarray}
where $\alpha \equiv 1-[4\Lambda +\ln (T/\nu_0)]/\zeta < 1$. Let us now
measure the energies in units of $\alpha \varepsilon_h$ and lengths in units
of $\alpha r_h $, where $\varepsilon_h \equiv T\zeta$. Again, the product $%
gr_h^3 \varepsilon_h$ can be estimated as $\beta/8$, and we obtain the
number of effective TRRs as $N_A \sim \mathcal{A}\alpha^7 S/r_h^2$, where
\begin{eqnarray}
&&\mathcal{A}=4\pi^2 (\beta/8)^2 \int \mathrm{d}\epsilon_l\int \! \mathrm{d}
\epsilon_s \int^{\epsilon_s}_0 \! \! \mathrm{d}\epsilon_d \int \! \eta_d^2\,
\mathrm{d}r_d \int\! \mathrm{d} \eta_s \int\! \mathrm{d}\eta_l  \notag \\
&& \times \Theta \left[\max(\epsilon_s,\epsilon_l) + \eta_d + \eta_s +
\eta_l) - 1\right]\approx 0.1 \, .  \label{NAR1}
\end{eqnarray}
Consequently, $\mathcal{G}/\mathcal{G}_n \sim \mathcal{A} \alpha^7 \ll 1$.
One concludes that the difference between the ``contact" resistances in
normal and superconducting states is dominated by the contribution of TRRs.
Since $\mathcal{G}_N \approx R_h^{-1}S/\mathcal{L}^2$,
\begin{equation}  \label{eq:mr1}
\delta R \equiv \mathcal{G}_n^{-1}-\mathcal{G}^{-1}\approx -\mathcal{G}%
^{-1}=-R_h\, (\mathcal{L}^2/S \mathcal{A} \alpha^7)\, .
\end{equation}
Note that the interface resistance is of the order of the resistance of HI
layer with the thickness $\sim \mathcal{L}/(\mathcal{A}\alpha^7) >> \mathcal{%
L}$. Since $\mathcal{L} \sim a \zeta^2$, one concludes that for $\zeta > 10$
the interface resistance can be comparable or even exceed the hopping
resistance if the thickness of the sample (or of the contact) is $\lesssim 10
$ $\mu$m.

The resistance estimated above can be experimentally measured as a
magnetoresistance in magnetic fields higher than the critical field for
superconductivity (similar effect for quasiparticle channel was studied in
\cite{Agrinskaya}).

We were implicitly assuming so far that the variable range hopping occurs
according to the Mott's law. This assumption certainly holds near the
interface where the Coulomb gap is screened by a superconductor. However in
the bulk of HI the Efros-Shklovskii (ES) law~\cite{CG,Shklovskii-Efros} can
become the dominant hopping mechanism. Then the value of $\zeta$ in the
connectivity criterion~(\ref{bind}) will be controlled by the Coulomb gap, $%
\zeta \to \zeta_{\text{ES}}=(\beta_1 e^2/\kappa a T)^{1/2}$, where $\beta_1$
is a numerical constant~\cite{CG,Shklovskii-Efros}. In this case, it turns
out that each bond of the ES backbone cluster finds some TRR ensuring charge
transfer. Thus in the limiting case of a weak tunneling interface barrier
the contact resistance will be the same for both  normal metal and
superconductor leads. This fact can be used to discriminate between Mott and
Efros-Shklovskii laws in the situation when it is difficult to do so from
temperature dependence.

Note that, in principle, the charge transfer involving double occupied
localized states is possible. However, such a process would require an
additional activation exponential factor, $\propto e^{-U/T}$, where $U$ is
the on-site correlation energy. One can also consider processes where a
double occupied center (so-called $D^-$-center) serves as an intermediate
state for the phonon-assisted two-electron tunneling. This channel is
unfavorable (at least in the case of a large interface barrier) because of
the above-mentioned exponential factor and a small pre-exponential factor
due to phonon-assisted tunneling. For the weak tunnel barriers the
conductance is controlled by ``typical" hopping sites. In this case $D^{-}$
channel is suppressed either by the additional tunneling exponential, $%
\propto e^{-4 r_h/a}$, or by a small probability to form a close triad of
hopping sites. Therefore the $D^{-}$ channel can be also neglected.

To conclude, we have developed a theory of the low-temperature charge
transfer between a superconductor and a hopping insulator and calculated the
interface resistance. This resistance is dominated by time reversal
reflection processes involving localized states in the insulator. It is the
time reversal reflection process that allows the low-temperature
measurements of hopping transport utilizing superconducting electrodes in
the experimental setups. In the Efros-Shklovskii VRH regime, the
corresponding interface resistance is small as compared to the bulk hopping
resistance and is nearly equal to the resistance at the interface between
the HI and normal metal. On the contrary, in the Mott hopping regime
(relevant, in particular, for 2D gated structures), the interface resistance
grows much larger and becomes commensurate (or even exceeds) to the bulk
hopping resistance. This effect is especially pronounced in the mesoscopic
samples. The contribution from the interface resistance can be detected by
application of the external magnetic field: the relatively weak magnetic
field will drive the superconductor into the normal state, but will not
affect the hopping transport eliminating thus time reversal reflection
process. This effect holds even in the case where the interface contribution
is less than the typical resistance of the hopping system itself.

\acknowledgments We are delighted to thank A. S. Ioselevich,
F.~W.~J.~Hekking, F. Pistolesi, and J. Bergli for discussion and
critical reading the manuscript. We are also grateful to L. I.
Glazman and B.~I.~Shklovskii for posing important questions. This
work was supported by the U. S. Department of Energy Office of
Science through contract No. W-31-109-ENG-38. A.A.Z. is grateful
to International Center for fundamental physics in Moscow and to
the Fund of noncommercial programs DYNASTY. 


\begin{thebibliography}{99}
\bibitem{end1} 
We assume that the on-site correlation (Hubburd) energy is sufficiently
large to prevent double occupation of the sites.
\end{thebibliography}
\end{document}